\documentclass[letter]{aa}

\usepackage{graphicx}
\usepackage{balance}
\usepackage{natbib}
\usepackage{setspace}
\usepackage{txfonts}
\usepackage{colortbl}
\usepackage{amssymb}
\usepackage{xspace}

\newcommand{\prot}{$P_\text{rot}$}

\newcommand{\teff}{$T_\text{eff}$}
\newcommand{\logg}{$\log\, g$}

\newcommand{\metal}{[Fe/H]}
\newcommand{\betHyi}{$\beta$\,Hyi\xspace}
\newcommand{\Aqr}{94\,Aqr\,Aa\xspace}
\newcommand{\Cyg}{16\,Cyg\,A\xspace}
\newcommand{\xray}{\mbox{X-ray}\xspace}

\definecolor{purple}{RGB}{102,0,204}
\definecolor{blue2}{RGB}{29,95,161}
\definecolor{orange}{RGB}{225,130,0}
\definecolor{green2}{RGB}{29,161,95}

\usepackage{hyperref,multirow}
\hypersetup{colorlinks,citecolor={blue2},linkcolor={blue2},urlcolor={blue2}}  

\begin{document}

\title{Magnetic braking and dynamo evolution of $\beta$ Hydri} 

\author{A. R. G. Santos\inst{1,2}
\and T. S. Metcalfe \inst{3}
\and O. Kochukhov \inst{4}
\and T. R. Ayres \inst{5}
\and R. Gafeira \inst{6,7}
\and T. L. Campante \inst{1,2}
}

\institute{Instituto de Astrof\'isica e Ci\^encias do Espa\c{c}o, Universidade do Porto, CAUP, Rua das Estrelas, PT4150-762 Porto, Portugal \\ 
\email{Angela.Santos@astro.up.pt}
\and Departamento de F\'isica e Astronomia, Faculdade de Ciências, Universidade do Porto, Rua do Campo Alegre 687, PT4169-007 Porto, Portugal
\and Center for Solar-Stellar Connections, White Dwarf Research Corporation, 9020 Brumm Trail, Golden, CO 80403, USA
\and Department of Physics and Astronomy, Uppsala University, Box 516, SE-75120 Uppsala, Sweden
\and Center for Astrophysics and Space Astronomy, 389 UCB, University of Colorado, Boulder, CO 80309, USA
\and Geophysical and Astronomical Observatory, Faculty of Science and Technology, University of Coimbra, Rua do Observat\'orio s/n, PT3040-004, Coimbra, Portugal
\and Instituto de Astrof\'isica e Ci\^encias do Espa\c{c}o, Department of Physics, University of Coimbra, Rua Larga, PT3040-004 Coimbra, Portugal
}
   
\date{\today}
   
\authorrunning{Santos, A. R. G., et al.}
\titlerunning{Magnetic braking and dynamo evolution}
 
\abstract
{The evolution of magnetic braking and dynamo processes in subgiant stars is essential 
for understanding how these stars lose angular momentum. We investigate the magnetic 
braking and dynamo evolution of \betHyi, a G-type subgiant, to test the hypothesis of 
weakened magnetic braking and the potential rejuvenation of large-scale magnetic fields. 
We analyze spectropolarimetric observations from HARPS (HARPSpol; polarimetric mode of High Accuracy Radial velocity Planet 
Searcher), and combine them with 
archival \xray data and asteroseismic properties from TESS (Transiting Exoplanet Survey Satellite) to estimate the current wind 
braking torque of \betHyi. Despite experiencing weakened magnetic braking during the 
second half of its main-sequence lifetime, our results indicate that \betHyi has regained 
significant magnetic activity and a large-scale magnetic field. This observation aligns 
with the ``born-again'' dynamo hypothesis. Furthermore, our estimated wind braking torque 
is considerably stronger than what would be expected for a star in the weakened magnetic 
braking regime. This suggests that subgiants with extended convective zones can 
temporarily re-establish large-scale dynamo action. These results provide critical 
constraints on stellar rotation models and improve our understanding of the interplay 
between magnetic field structure, stellar activity cycles, and angular momentum evolution 
in old solar-type stars.}
  

\maketitle

\section{Introduction}\label{sec1}

Subgiants can provide some of the most stringent tests of weakened magnetic braking 
\citep[WMB;][]{vanSaders2016} in old solar-type stars. During the early phases of stellar 
evolution, rotation slows as the magnetized stellar wind gradually sheds angular 
momentum, a process known as magnetic braking \citep{Weber1967, Skumanich1972, 
Kawaler1988}. According to the WMB hypothesis, when rotation becomes sufficiently slow, 
the stellar dynamo can no longer organize magnetic fields on the largest spatial 
scales---inhibiting further angular momentum loss \citep{Reville2015, Garraffo2016} and 
keeping the rotation rate approximately constant during the second half of the 
main-sequence lifetime. The rotation periods predicted by WMB and standard rotational 
evolution models subsequently diverge, reaching a maximum fractional difference on the 
subgiant branch \citep{vanSaders2019}. Observations of the G8 subgiant \Aqr provided an 
early test of WMB, which successfully explained a rotation period about half as long as 
that predicted by standard models \citep{Metcalfe2020}.

If the onset of WMB corresponds to a transition in the underlying dynamo, then there 
should be observable signatures in the evolution of stellar activity cycles. This idea 
was first proposed by \cite{Metcalfe2017}, who suggested a gradual lengthening and 
weakening of activity cycles during the second half of main-sequence lifetimes, 
ultimately leading to the ``flat activity'' stars in the Mount Wilson survey 
\citep{Wilson1978, Baliunas1995}. With few exceptions, most of the subgiants with 
long-term stellar activity measurements show flat activity. The exceptions are either 
stars that originated above the \cite{Kraft1967} break and only developed a substantial 
convection zone as a subgiant \citep[e.g., HD\,81809;][]{Egeland2018}, or stars slightly 
more massive than the Sun that rejuvenated their activity cycles with a growing 
convection zone on the subgiant branch---the ``born-again'' dynamo scenario described by 
\cite{Metcalfe2020, Metcalfe2024b}. Both \Aqr and \betHyi are in this latter category of 
exceptions, which is only possible with WMB \citep[see][their Fig.~5]{Metcalfe2024b}.

A key question is whether the born-again dynamo phase is accompanied by a return of the 
large-scale magnetic field, and the ensuing higher rate of angular momentum loss. To 
address this question, we estimate the current wind braking torque of \betHyi following 
the prescription of \cite{FinleyMatt2018}. In Section~\ref{sec2} we assemble the required 
inputs, including spectropolarimetry to constrain the large-scale magnetic field 
strength, \xray measurements to constrain the mass-loss rate from the empirical relation 
of \cite{Wood2021}, the rotation period from the Transiting Exoplanet Survey Satellite 
\citep[TESS;][]{Ricker2014} and the asteroseismic stellar mass and radius 
\citep{Metcalfe2024b}. In Section~\ref{sec3} we use these inputs to estimate the wind 
braking torque and we evaluate the results in the context of similar observations of 
several solar analogs and slightly hotter stars. Finally, in Section~\ref{sec4} we 
discuss our results, concluding that the onset of WMB may coincide with a threshold for 
the decay of large-scale dynamo action, which can be temporarily reversed in the 
born-again phase.

\section{\betHyi observations and properties}\label{sec2}

\subsection{Spectropolarimetry and large-scale field}\label{sec2.1} 

We observed \betHyi on July 14, 2024 with the High Accuracy Radial velocity Planet 
Searcher \citep[HARPS;][]{Mayor2003}, using the polarimetric mode 
\citep[HARPSpol;][]{Piskunov2011, Snik2011}. This instrument configuration allowed us to 
obtain intensity (Stokes~$I$) and circular polarization (Stokes~$V$) spectra 
simultaneously, with a resolving power of 110\,000 and a wavelength coverage from 379 to 
691~nm, except for a gap in the 526--534~nm region. The target was observed for $\sim$5 
h, yielding 49 independent observations. Each observation consisted of four 30~s 
sub-exposures, between which the quarter-wave plate in the HARPSpol polarimeter was 
rotated to exchange positions of the orthogonally polarized beams on the detector. This 
procedure followed the standard spatio-temporal spectropolarimetric modulation scheme 
widely used in high-resolution polarimetry \citep{Donati1997,Bagnulo2009}. The spectra 
were reduced using the \textsc{reduce} package \citep{Piskunov2002}, following the 
procedure described in \citet{Rusomarov2013}. The resulting data are characterized by a 
signal-to-noise ratio of $\sim$200 per pixel at $\lambda=550$~nm.

As is typical for most cool stars, observational noise precludes the detection of 
polarization signatures in individual spectral lines of \betHyi. To overcome this, we 
calculated least-squares deconvolved \citep[LSD;][]{Donati1997, Kochukhov2010} Stokes~$I$ 
and $V$ profiles by combining numerous spectral lines. The LSD line mask was obtained 
from the VALD database \citep{Ryabchikova2015}, using $T_{\rm eff}=5750$~K and $\log 
g=4.0${, close to the literature values of atmospheric parameters of \betHyi. Retaining 
spectral lines with the central depth exceeding 10\% of the continuum and excluding 
regions affected by broad stellar lines and telluric features resulted in about 4800 
metal lines suitable for LSD. The line-addition procedure was applied to the 49 
observations individually and then the resulting profiles were combined, achieving a 
polarimetric sensitivity of $\approx8\times10^{-6}$ per 0.8~km\,s$^{-1}$ velocity bin in 
Stokes~$V$. The mean circular polarization profile of \betHyi, illustrated in 
Fig.~\ref{fig1}, shows a clear detection of a Zeeman signature. The false alarm 
probability of this detection is $<$\,10$^{-6}$ indicating a definite detection according 
to the conventions in high-resolution spectropolarimetry \citep{Donati1997}. This 
Stokes~$V$ profile corresponds to a mean longitudinal magnetic field of $\langle B_{\rm 
z} \rangle=-0.298\pm0.086$~G.

To characterize the global magnetic field of \betHyi, we fit the observed LSD Stokes~$V$ 
profile with the modeling procedure from \citet{Metcalfe2019}. This modeling approach, 
based on the assumption of an axisymmetric dipolar field morphology, requires adopting a 
stellar inclination angle, $i$. We follow the procedure by \citet{Bowler2023} to 
constrain the $i$ posterior distribution, knowing the projected rotational velocity, 
rotation period, and stellar radius. We find $i=50^{\circ+21}_{\,\,\,-14}$. The best fit 
to the polarization profile, assuming an axisymmetric dipole (in red), returned a polar 
magnetic field strength of $B_{\rm d}=-0.64$ G. However, as seen in Fig.~\ref{fig1}, the 
fit to the data is relatively poor with a reduced $\chi^2$ of 3.0. The likely cause for 
this discrepancy is the presence of dominant non-axisymmetric field components at the 
time of our observation of \betHyi. To account for this more complex global field 
geometry, we made use of the \textsc{InversLSD} code \citep{Kochukhov2014}, running it in 
a highly restrictive mode with only dipolar poloidal components allowed in the harmonic 
expansion ($\ell_{\rm max}=1$, $\beta=\alpha$, and $\gamma=0$ according to the harmonic 
coefficient definitions in \citealt{Kochukhov2014}). This approach leads to an 
improvement of the fit, with $\chi^2=1.3$. In this case, the best-fit parameters 
correspond to a dipolar field strength of $B_\text{d}=2.13$~G and an obliquity angle of 
$\beta=87.3^\circ$.

\begin{figure}[t!]
\centering
\includegraphics[width=\linewidth]{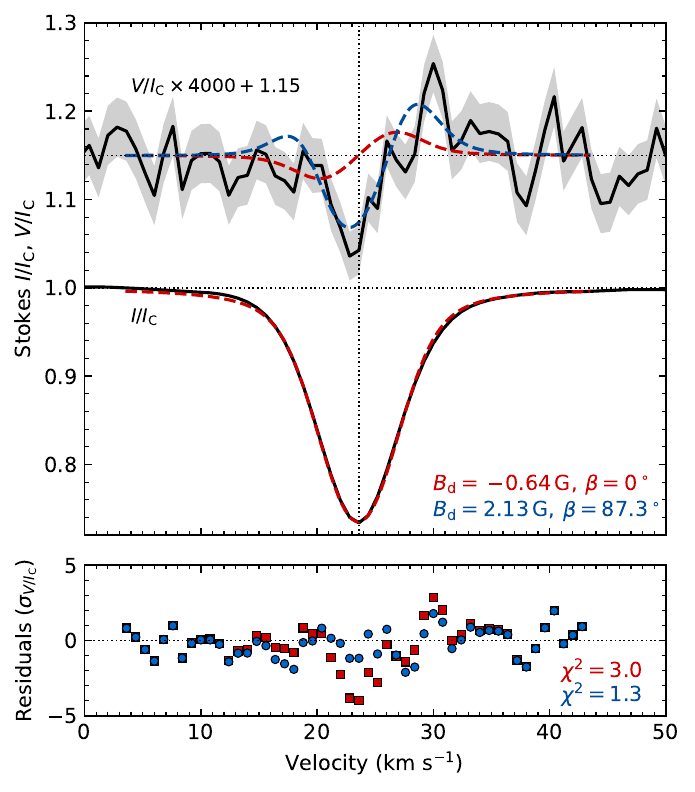}
\caption{LSD Stokes~$V$ and $I$ profiles for \betHyi derived from HARPSpol 
observations (top). For illustration purposes, Stokes~$V$ is scaled and shifted 
vertically in relation to Stokes~$I$. The black solid line shows the mean observed 
profile, with the respective uncertainty indicated by the gray-shaded region. The 
vertical dotted line marks the line center. The red and blue lines indicate the best-fit 
models obtained for a dipole field, considering an axisymmetric and an inclined field, 
respectively. The red squares and blue circles show the corresponding residuals for the Stokes $V$ profile (bottom). 
The fitting parameters and $\chi^2$ values are annotated within the panels.\label{fig1}} 
\end{figure} 

\subsection{X-ray data and mass-loss rate}\label{sec2.2}

\betHyi is an active subgiant star known to harbor a magnetic activity cycle 
\citep{Dravins1993,Metcalfe2007}. The cycle period ($P_\text{cyc}$) is estimated to be 
approximately 12~yr, based on the Mg~\textsc{ii} activity proxy \citep{Metcalfe2007} with 
data from the International Ultraviolet Explorer \citep[IUE; ][]{Boggess1978a, 
Boggess1978b}. \citet{Dravins1993} found \betHyi to have a lower mean activity level than 
the Sun. However, compared to the weaker Solar Cycle 24 discussed below, \betHyi has a 
higher mean activity level.

Fig.~\ref{fig2} shows the \xray luminosity of \betHyi (circles), relative to its 
bolometric luminosity $L_\text{BOL}$. The gray line depicts the 2008–2019 \xray 
modulation of Solar Cycle 24 \citep{Ayres2021}, as derived by smoothing daily values in 
the 0.1–2.4 keV bandpass with an 81-day running mean (three solar synodic rotations). The 
single-cycle template was then replicated over several 11-year intervals, and scaled in 
period (using $P_\text{cyc}=12$~yr) and amplitude to match the \xray variability of 
\betHyi seen during the ROSAT \citep[R\"ontegen Satellit;][]{Trumper1982} era 1990–1998. 
The latter was based on archival \xray count rates from multiple sources: ROSAT All Sky 
Survey (yellow); ROSAT Position Sensitive Proportional Counters catalog (orange); and 
ROSAT High Resolution Imager catalog (red). For comparison, the white squares show the 
IUE Mg~\textsc{ii} fluxes \citep{Metcalfe2007} shifted in the mean and stretched in 
amplitude to match the apparent \xray high and low values. The blue and green circles 
show the recent \xray luminosities from XMM-Newton EPIC \citep[\xray Multi-Mirror; The 
European Photon Imaging Camera;][]{Jansen2001}, and Chandra ACIS \citep[Advanced CCD 
(Charge-Couple Device) Imaging Spectrometer; ][]{Weisskopf2000,Weisskopf2002}, both being 
custom-processed \citep[details in][]{Ayres2025}. In all cases, the \xray count rates 
were converted to fluxes using model-derived energy conversion factors \citep{Ayres2025}, which also depend on the instrument response function.

\begin{figure}[t!]
\centering
\includegraphics[width=\linewidth]{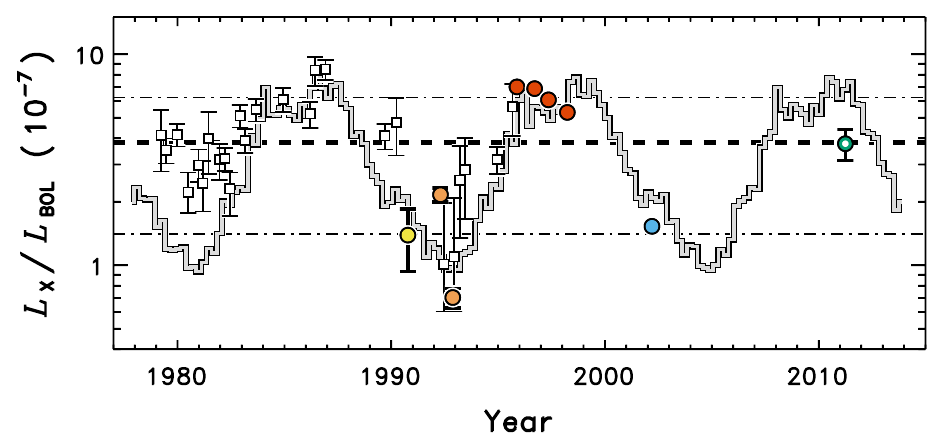}
\caption{\xray to bolometric luminosity ratio of \betHyi (circles). The light gray curve 
mimics the solar \xray variation over Cycle~24, replicated in time and stretched to a 
period of 12 yr, then adjusted to the apparent \xray modulation amplitude of \betHyi. 
Archival ROSAT fluxes from various catalogs are shown as the yellow, orange, and red 
circles, XMM-Newton as the blue circle, and Chandra as the green open circle. Horizontal 
dashed and dot-dashed lines mark the mean normalized \xray luminosity of \betHyi and $\pm 
1\sigma$ uncertainties. White squares represent the Mg~\textsc{ii} time-series scaled to 
match the amplitude of the \xray variation.\label{fig2}}
\end{figure} 

The combined data reproduce the peaks and valleys of the \xray variability of \betHyi 
tolerably well, supporting the 12~yr cycle period originally derived solely from 
Mg~\textsc{ii} \citep{Metcalfe2007}. The corresponding mean \xray luminosity is $\langle 
L_\text{X}\rangle = 5.1\pm3.1\times 10^{27} \text{erg\,s}^{-1}$. The uncertainty, taken 
as the standard deviation of the \xray measurements, is a good proxy for the amplitude of 
the high-energy variation over the activity cycle. To estimate the mass-loss rate 
$\dot{M}$ of \betHyi, we use the empirical relation $\dot{M} \propto 
F_\mathrm{X}^{0.77\pm0.04}$ from \citet{Wood2021}, where $F_\text{X}$ is the \xray 
surface flux, computed from $\langle L_\text{X}\rangle$ and the asteroseismic radius. We 
obtain a current mass-loss rate of $\dot{M} = 0.80^{+0.40}_{-0.44}\, \dot{M}_\odot$.

\begin{table}[t!]
\centering
\caption{Properties of the solar-type subgiant \betHyi.\label{tab1}}
\begin{tabular}{lcc}
\hline\hline
& \betHyi & Source \\\hline
\teff\ (K) & $5872 \pm 74$ & (1) \\
\metal\ (dex) & $-0.10 \pm 0.09$ & (2) \\
\logg\ (dex) & $3.84 \pm 0.08$ & (2)\\
$v\sin i$ (km s$^{-1}$) & $2.7 \pm 0.6$ & (2)\\
B$-$V (mag) & 0.618 & (3)\\
$\log R'_\text{HK}$  (dex) & $-4.996 \pm 0.047$ & (3) \\
$P_\text{cyc}$ (yr) & $12.0_{-1.7}^{+3.0}$ & (4) \\
$i$ ($^\circ$) & $50_{-14}^{+21}$ & (5) \\
$|B_\text{d}|$ (G) & 2.13 & (5) \\
$\langle L_\text{X}\rangle$ ($10^{27}$~erg~s$^{-1}$) & $5.1 \pm 3.1$ & (5) \\
Mass-loss rate ($\dot{M}_\odot$) & $0.80^{+0.40}_{-0.44}$  & (5) \\
\prot\ (days) & $23.0 \pm 2.8$ & (6) \\
Ro/Ro$_\odot$ & $0.959 \pm 0.117$ & (6)\\
Mass ($M_\odot$) & $1.127 \pm 0.054$ & (6) \\
Radius ($R_\odot$) & $1.840 \pm 0.032$ & (6) \\
$L_\text{BOL}$ ($L_\odot$) & $3.45 \pm 0.10$ & (6) \\
Age (Gyr) & $6.26 \pm 0.57$ & (6) \\\hline
Torque ($10^{30}$~erg) & $3.51^{+1.78}_{-1.55}$  & (7) \\\hline\hline
\end{tabular}
\tablebib{(1)~\citet{North2007}; (2)~\citet{Bruntt2010}; (3)~\citet{Henry1996}; 
(4)~\citet{Metcalfe2007}; (5)~Section~\ref{sec2}; (6)~\citet{Metcalfe2024b}; 
(7)~Section~\ref{sec3}.}
\end{table}

\section{Wind braking torque}\label{sec3}

We now have everything required to estimate the wind braking torque for \betHyi, 
following the prescription of 
\cite{FinleyMatt2018}\footnote{\url{https://github.com/travismetcalfe/FinleyMatt2018}}. 
Combining the large-scale magnetic field strength from spectropolarimetry in 
Section~\ref{sec2.1}, the mass-loss rate from the empirical relation of \cite{Wood2021} 
in Section~\ref{sec2.2}, and the rotation period as well as the asteroseismic mass and 
radius from TESS photometry \citep{Metcalfe2024b}, we calculate a wind braking torque of 
$3.51^{+1.78}_{-1.55} \times 10^{30}$~erg (see Table~\ref{tab1}). The uncertainties are 
defined by simultaneously shifting all of the inputs to their $\pm 1\sigma$ values to 
minimize or maximize the torque.

In Fig.~\ref{fig3}, we compare \betHyi with similarly estimated wind braking torques for 
two late F-type stars \citep{Metcalfe2021}, and five solar analogs \citep{Metcalfe2022, 
Metcalfe2024a}. Rossby numbers ($\mathrm{Ro} \equiv P_\mathrm{rot}/\tau_c$) were 
calculated using the observed rotation periods and the asteroseismic calibration of 
convective turnover time from \cite{Corsaro2021}, normalized to the solar value on this 
scale (Ro$_\odot=0.496$). The wind braking torque is normalized to the value for 
HD\,76151 ($4.17 \times 10^{30}$~erg) to facilitate a comparison with theoretical models. 
The gray shaded area represents an empirical constraint on the critical Rossby number for 
the onset of WMB \citep[$\mathrm{Ro_{crit}}/\mathrm{Ro_\odot}=0.92\pm0.01$;][]{Metcalfe2024a}, 
and the dotted yellow line shows the evolution of the torque for HD\,76151 from a 
standard spin-down model \citep{vanSaders2013}. For \betHyi, the horizontal error bar is 
dominated by the rotation period uncertainty, while the vertical error bar is dominated by 
variation of the \xray surface flux through the activity cycle.

\begin{figure}
\centering
\includegraphics[width=\linewidth]{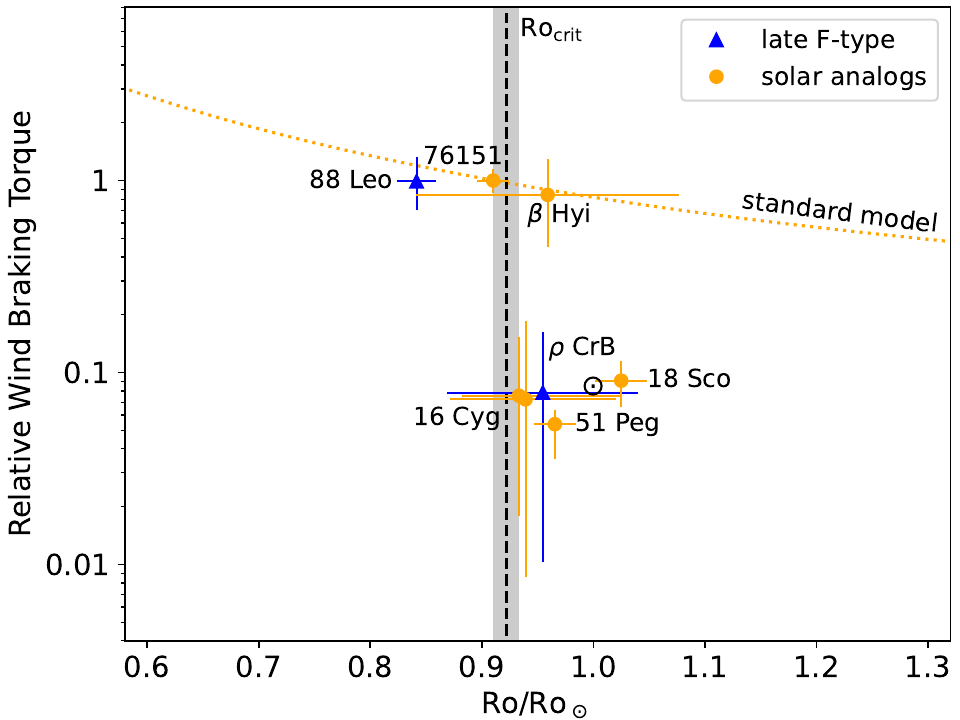}
\caption{Estimated wind braking torque relative to HD\,76151 as a function of Rossby 
number normalized to the solar value. Points are grouped by spectral type, as indicated 
in the legend. The gray shaded area represents an empirical constraint on the critical 
Rossby number for the onset of WMB \citep[$\mathrm{Ro_{crit}}/\mathrm{Ro_\odot}=
0.92\pm0.01$;][]{Metcalfe2024a}. The solar wind braking torque was taken from \cite{Finley2018_massloss}. \label{fig3}}
\end{figure}

The Rossby number of \betHyi places it between $\mathrm{Ro_{crit}}$ and the solar value 
\citep[cf.][their Fig.~5]{Metcalfe2024b}, where the wind braking torque decreases 
abruptly by an order of magnitude in middle-aged dwarfs. Compared to other solar analogs 
in the WMB regime such as \Cyg, the wind braking torque for \betHyi is nearly an order of 
magnitude stronger. To assess the underlying sources of this stronger torque, we can 
change one parameter at a time between the fiducial models for \betHyi and \Cyg 
\citep{Metcalfe2022}. Small decreases in torque would be expected from the longer 
rotation period ($-11\%$), the lower estimated mass-loss rate ($-7\%$), and the slightly 
higher stellar mass ($-1.1\%$) of \betHyi. However, these differences are overwhelmed by 
the increase in torque from the much stronger magnetic field ($+280\%$) and the larger 
stellar radius ($+260\%$). Together, these parameters increase the surface magnetic flux 
and yield a larger Alfv\'en radius (lever-arm) for a given mass-loss rate in the 
simulations of \cite{FinleyMatt2018}.

\section{Discussion and Conclusions}\label{sec4} 

The evolution of stellar magnetism is closely linked to changes in the rotation and 
convective properties of stars. During the main-sequence, the Ro evolution is dictated by 
the \prot\ evolution, which generally increases with age. In the WMB scenario, stars 
gradually lose their magnetic cycles, while the \prot\ evolution due to magnetic braking 
almost ceases. In the subgiant phase, as the star expands, the convective turnover time 
grows longer, dominating over the \prot\ evolution and contributing to a decreasing Ro. 
This can eventually bring Ro back below $\mathrm{Ro_{crit}}$. As a result, the 
large-scale dynamo can be rejuvenated in the subgiant phase, allowing stars to sustain 
activity cycles once more---a phenomenon known as the born-again dynamo 
\citep{Metcalfe2020, Metcalfe2024b}. Therefore, subgiant stars are valuable for testing 
stellar evolution scenarios, particularly those related to stellar magnetism. Our 
analysis focused on the subgiant \betHyi, which harbors a magnetic activity cycle more 
typical of a younger solar-like dynamo, contrary to expectations and supporting the 
born-again dynamo scenario. Combining HARPSpol magnetic field constraints, the \xray 
surface flux, and other stellar properties such as the rotation period, mass, and radius, 
we estimated the current mass-loss rate and wind braking torque of \betHyi. The latter 
exceeds that of the Sun by nearly an order of magnitude, suggesting the return of a 
stronger large-scale field coinciding with the born-again dynamo phase. Despite the 
current wind braking torque of \betHyi, \cite{Metcalfe2024b} demonstrated that standard 
spin-down models cannot match its observed rotation period, and Ro never returns below 
$\mathrm{Ro_{crit}}$ in the subsequent evolution. By contrast, WMB models do match the 
observed rotation period and also bring Ro close to $\mathrm{Ro_{crit}}$ at its current 
asteroseismic age. In other words, \betHyi appears to have experienced WMB near the 
middle of its main-sequence lifetime, after reaching $\mathrm{Ro_{crit}}$ earlier in its 
evolution.

If the value of $\mathrm{Ro_{crit}}$ for the onset of WMB is also the critical value for 
the efficient organization of large-scale fields, then the activity cycle in \betHyi may 
represent a subcritical dynamo \citep{Tripathi2021}. \cite{DurneyLatour1978} suggested 
that the dynamo number $D \propto \mathrm{Ro}^{-2}$, and low values of $D$ are 
insufficient to organize large-scale fields until reaching a critical value $D_{\rm 
crit}$. However, due to hysteresis in the system, models that are initialized with 
magnetism can continue to organize large-scale field even when $D < D_{\rm crit}$, 
resulting in a ``subcritical dynamo'' \citep{Vashishth2021}. This mirrors stellar 
evolution, where stars begin with higher magnetic activity levels and low Ro, gradually 
becoming less active over time while Ro increases. This leads to the possibility that 
$\mathrm{Ro_{crit}}$ may correspond to $D_{\rm crit}$, and dynamos that operate with 
$\mathrm{Ro} > \mathrm{Ro_{crit}}$ are subcritical. The rejuvenation of large-scale field 
and magnetic braking in \betHyi above $\mathrm{Ro_{crit}}$ supports the idea that its 
activity cycle may be driven by a subcritical dynamo. However, the rotation period 
uncertainty does not currently exclude the possibility that $\mathrm{Ro} < 
\mathrm{Ro_{crit}}$.

Future observations of \betHyi may refine our estimate of the wind braking torque. We 
used a single spectropolarimetric measurement to constrain the large-scale magnetic field 
by modeling the Stokes~$V$ profile with an inclined dipole. Observations spanning a 
complete rotation would allow for Zeeman-Doppler Imaging, providing the detailed 
morphology of the magnetic field, which may be more complex. We estimated the mass-loss 
rate from the \xray surface flux using the empirical relation of \cite{Wood2021}. 
However, direct inferences of the mass-loss rate from Ly$\alpha$ observations can deviate 
substantially from this relation, particularly for subgiants (e.g., $\delta$\,Pav and 
$\delta$\,Eri). Ly$\alpha$ measurements of \betHyi from the Hubble Space Telescope are 
currently scheduled for April 2025 (HST-GO-17793, PI:~B.\,Wood), so a direct inference of 
the mass-loss rate may soon be possible. The measured rotation period still has a large 
uncertainty, but the TESS mission will observe \betHyi again for two consecutive sectors 
(54 days) in July/August 2025, providing an opportunity to confirm or refine the current 
measurement. These new constraints on the wind braking torque should help clarify whether 
the activity cycle in \betHyi truly represents a subcritical dynamo, or if long-term 
UV/\xray measurements of other stars with apparently constant activity in Ca~HK (e.g., 
$\rho$\,CrB and \Cyg \& B) might also reveal cycling behavior. As one of only two low-mass 
subgiants currently known to exhibit an activity cycle, \betHyi is a crucial benchmark for 
testing and refining models of magnetic braking and dynamo evolution in old solar-type stars.

\begin{acknowledgements}
This work is based on observations collected at the European Organisation for 
Astronomical Research in the Southern Hemisphere under ESO programme 113.26SZ.001. It was 
supported by Funda\c{c}\~ao para a Ci\^encia e a Tecnologia (FCT) through the research 
grants UIDB/04434/2020 (DOI: 10.54499/UIDB/04434/2020), UIDP/04434/2020.FCT (DOI: 
10.54499/UIDP/04434/2020) and 2022.03993.PTDC (DOI: 10.54499/2022.03993.PTDC). This 
research was also supported by the International Space Science Institute (ISSI) in Bern, 
through ISSI International Team project 24-629 (Multi-scale variability in solar and 
stellar magnetic cycles). A.R.G.S.\ acknowledges the support from the FCT through the work 
contract No. 2020.02480.CEECIND/CP1631/CT0001 (DOI: 
10.54499/2020.02480.CEECIND/CP1631/CT0001). T.S.M.\ acknowledges support from NSF grant 
AST-2205919 and NASA grant 80NSSC22K0475. O.K.\ acknowledges support by the Swedish 
Research Council (grant agreement no. 2023-03667) and the Swedish National Space Agency. 
T.L.C.\ is supported by FCT in the form of a work contract 
(2023.08117.CEECIND/CP2839/CT0004; DOI: 10.54499/2023.08117.CEECIND/CP2839/CT0004).
\end{acknowledgements}

\bibliographystyle{aa}
\bibliography{betaHyi}

\end{document}